# A high fidelity approximation of radial polarization conversion

*P.B. Phua[1,2], W.J. Lai[2], Yuan Liang Lim[1]*

[1] *DSO National Laboratories, 20 Science Park Drive, Singapore 118230.*
[2] *Nanyang Technological University, 1 Nanyang Walk, Blk 5 Level 3, Singapore 637616.*

**Abstract:**

We report a high fidelity (>90%) approximation of radial polarization conversion using a birefringent lens. It offers the advantages of low-cost, easy fabrication, alignment robustness and high laser power handling capability.

# A high fidelity approximation of radial polarization conversion

P.B. Phua[1,2], W.J. Lai[2], Yuan Liang Lim[1]

[1] DSO National Laboratories, 20 Science Park Drive, Singapore 118230.
[2] Nanyang Technological University, 1 Nanyang Walk, Blk 5 Level 3, Singapore 637616.

External radial polarization converter that generates an approximation of radial polarization, has been reported in [1,2]. In this paper, we report an attractive method that offers low-cost, easy fabrication, robustness, high fidelity approximation and high power laser handling capability. This scheme approximates the radial polarized light using a custom designed birefringent lens whose crystallographic optics axis is aligned to the optical axis of the lens. Thick birefringent lens with such crystal's orientation has been proposed in [2]. In their experiment, the radially and tangentially polarized components of a circularly polarized input beam, experience different focusing power which leads to different focal points along the optical axis. This bi-focusing effect allows one to separate out the radial polarized light by placing a pin-hole at the appropriate focal point. However, the efficiency to obtain the radial polarized power from such scheme is capped at 50%.

Our new approach of using birefringent lens does not rely on bi-focusing. In fact, we use thin birefringent lens that has negligible bi-focusing effect. We design its radius of curvature and center thickness of the lens so as to create the appropriate polarization transformation to approximate the radial polarization. This approach can now increase the efficiency of converting circularly polarized light to radial polarization to more than 90%.

Our plano-convex birefringent lens is fabricated using crystalline quartz whose crystallographic optics axis is aligned to the optical axis of the lens. Collimated rays of circularly polarized light entering the convex surface of the lens are bent into various angles, $\theta(r)$ where $r$ is the distance from the lens optical axis as shown in Fig. 1. As a result, the birefringence $\Delta n$ and the crystal length $l$ traversed, varying with $r$, yield the ray a polarization retardation angle of $\phi(r) = \frac{2\pi}{\lambda}\Delta n(r) l(r)$ as shown in Fig. 2. The radius of curvature and center thickness of the lens used is 6 cm and 2.2 mm respectively, designed for an input wavelength of 532 nm. With the slow polarization eigen-axis of this birefringent lens in the radial direction, the output polarization after passing through the birefringent lens, followed by a 45° optical activity quartz rotator, is approximately radial. Fig. 3 shows the computed percentage of power in the radial polarization, $P$, as a function of $r$. It reaches 100% when $\phi$ equals to $\pi/2$ radian. Clearly, at the beam center (i.e. $r = 0$), $P$ is 50% because this ray propagates along the crystal's optics axis and does not experience any birefringence. Thus the output beam is still circularly polarized light at the beam center.

If we define the radial polarization purity of the beam as the weighted average of $P(r)$ across the beam and the weighting function is the normalized beam intensity, the radial polarization purity can be computed to be > 94% for an input beam of Laguerre-Gaussian $LG_{01}$ mode and >85% for a Gaussian mode. The radial polarization purity of $LG_{01}$ mode is expected to be higher than Gaussian mode since it has more power concentrated around the peak of $P$. Figure 4 shows the approximated radially polarized beams when a Gaussian mode is used as an input beam. They were experimentally measured with and without a polarizer. The observed 'dumbbell' rotating with the transmitting axis of the polarizer, demonstrates the approximated radially polarized beam.

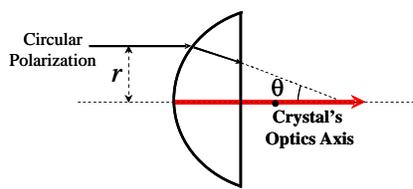

**Fig. 1**

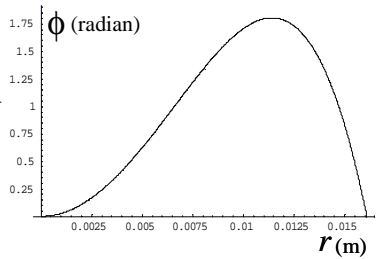

**Fig. 2**

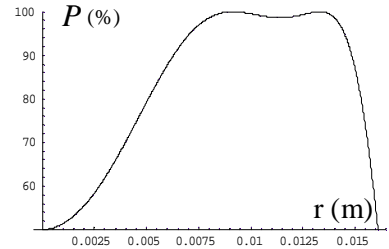

**Fig. 3**

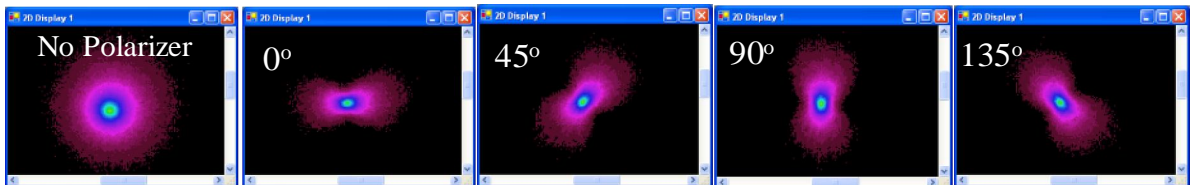

**Fig. 4**